\documentclass[aps,prb,preprint,groupedaddress,showpacs,floatfix]{revtex4}

\usepackage{graphicx}

\newcommand{\LiTiO}{LiTi$_2$O$_4$}

\newcommand{\LiAlTiO}{LiAl$_y$Ti$_{2-y}$O$_4$}

\begin{document}

\title{Electronic properties of\\ disordered corner-sharing tetrahedral 
lattices}

\author{F. Fazileh}
\altaffiliation{Present Address: Department of Physics, University of Windsor, 
Windsor ON N9B 3P4 Canada.}
\affiliation{Department of Physics, Queen's University, Kingston ON K7L 3N6 
Canada}
\author{X. Chen}
\affiliation{Department of Physics, Queen's University, Kingston ON K7L 3N6 
Canada}
\author{R. J. Gooding}
\affiliation{Department of Physics, Queen's University, Kingston ON K7L 3N6 
Canada}
\author{K. Tabunshchyk}
\altaffiliation{Present Address: NINT and the Department of Engineering 
Physics, University of Alberta, Edmonton AB T6G 2V4 Canada.}
\affiliation{Department of Physics, Queen's University, Kingston ON K7L 3N6 
Canada}

\date{\today}

\begin{abstract}
We have examined the behaviour of noninteracting electrons moving on a 
corner-sharing tetrahedral 
lattice into which we introduce a uniform (box) distribution, of width $W$, 
of random on-site energies. 
We have used both the relative localization length and the spectral rigidity 
to analyze the nature of 
the eigenstates, and have determined both the mobility edge trajectories as 
a function of $W$,
and the critical disorder, $W_c$, beyond which all states are localized. 
We find (i) that the mobility edge 
trajectories (energies $E_c$ versus disorder $W$) are qualitatively 
different from those found for a 
simple cubic lattice, and (ii) that the spectral rigidity is scale invariant 
at $W_c$ and thus provides 
a reliable method of estimating this quantity -- we find $W_c/t = 14.5$.  
We then discuss our results in the 
context of the metal-to-insulator transition undergone by 
LiAl$_y$Ti$_{2-y}$O$_4$~in a 
quantum site percolation model that also includes the above-mentioned Anderson 
disorder. We show that the effects of an inhomogeneous distribution of on-site 
energies produced by the Al impurity potentials are small compared to
those produced by quantum site percolation, at least in the determination 
of the doping concentration at which the metal-to-insulator transition is 
predicted to occur.
\end{abstract}


\maketitle

\section{Introduction}

The properties of the corner-sharing tetrahedral lattice (hereafter denoted
by CSTL) are of interest in a wide variety of physical systems.\cite{CJP01} 
This lattice is a non-bipartite, and thus frustrated, three dimensional 
structure 
that is the conducting path of electrons in many interesting systems. This 
network can be derived, \textit{e.g.}, from the diamond structure (which is 
a bipartite lattice) by placing a site at each bond midpoint.\cite{PWA56} 
One interesting consequence from this construction is that, \textit{e.g.}, 
the critical 
(classical) site percolation threshold of pyrochlore is the same as the bond 
percolation threshold of diamond. Also, this structure is a sublattice of 
many compounds, including pyrochlores and spinels. In the A$_2$B$_2$O$_7$ 
pyrochlore structure, both the A and B sublattices can be realized as a 
network of corner-sharing tetrahedra, which is nowadays often referred to as 
the pyrochlore lattice. Examples of the interesting (quantum) magnetic 
properties 
and superconducting behaviour of several pyrochlores are discussed in the
literature.\cite{pyroM,pyroSC} Indeed, there exist many experimental results 
which seem to suggest that the geometrical frustration of the CSTL
in these compounds is responsible for their peculiar properties, and 
it has been argued that geometrical frustration tends to amplify the 
correlation and quantum effects.\cite{CJP01} 

The initial motivation for our work has to do with the electronic properties 
of the 
normal spinel AB$_2$O$_4$,\cite{Barth32} in which the B sublattice forms
a corner-sharing tetrahedral network. There are a large number of well known 
spinels, 
some with exotic magnetic\cite{PWA56} and electronic 
properties.\cite{Goodenough60}
Further, the recent discovery\cite{dcj97} of the first $d$-electron heavy 
fermion compound 
LiV$_2$O$_4$ has generated substantial interest in such electronic systems, 
which has subsequently directed us to a study\cite{Fazileh04,Fazileh05}
of the simpler but as yet not understood properties of 
LiTi$_2$O$_4$,\cite{dcj76} the latter of which has been suggested to be 
related to the high-T$_c$ superconductors owing to strong electronic 
correlations.\cite{muller96} 

LiTi$_2$O$_4$ undergoes a metal-to-insulator transition when excess Li is 
doped onto the Ti (corner-sharing tetrahedral) sublattice, or when Al 
(or Cr, which is not discussed in this paper) is substituted for Ti on the 
same sublattice, and due to the large difference in on-site energies 
(\textit{e.g.}, doped Li versus Ti) this system is well approximated 
by a quantum site percolation model in which the removed sites are those Ti 
sites onto which either an excess Li or substituting Al ion are added.
Recently, two of the present authors and Johnston\cite{Fazileh04} have 
examined the possibility that such
a model accounts for the metal-to-insulator transition, and found that physics
beyond that contained in a quantum site percolation model will be required. 
One example candidate for this ``extra physics" can be understood as follows: 
the excess Li or substitutional Al impurity ions will generate a distribution 
of on-site energies around them due to their impurity potentials, so that in 
addition
to their presence eliminating those sites from the conducting path, their
effect on the neighbouring sites should also be included. As suggested by 
Anderson,\cite{PWA56} 
the role played by such physics is approximated by a random set of on-site 
energies. Here we focus on one simple variant of such random energies, 
\textit{viz.} that given by a uniform (box) distribution, and at the end of 
this paper we discuss the effect 
of such randomness on this metal-to-insulator transition.

We stress, however, that in any systematic study of the physics involving
itinerant electrons encountered on a disordered CSTL could require the 
information contained in our paper. For example, in addition to our own
work,\cite{Fazileh05,FazilehPhD} there are now several papers by 
Fujimoto\cite{fuji01,fuji02a,fuji02b,fuji03} and others\cite{chen02,imai02} 
on the physics of correlated electrons on this lattice, and if such work is 
extended to disordered systems, possessing an understanding of the disordered 
but uncorrelated electrons would be beneficial.

Our paper is organized as follows. In \S II we present the model that we study,
and in \S III we state the numerical procedures that we use, and the 
statistical quantities that we evaluate. In \S IV we give our comprehensive 
numerical results for
the mobility edge trajectories and the critical disorder, and \S V we present
our results for the quantum-site percolation model plus Anderson disorder that
is meant to mimic the (non-interacting) model of the metal-to-insulator 
transition in \LiTiO. Finally, in \S VI we state our conclusions.

\section{Model}

In an effort to better understand the physics of electrons moving on
this structure, we calculate numerically the mobility edge trajectories 
in the energy-disorder plane of a CSTL with a tight-binding Hamiltonian with 
a uniform (box) distribution model of disorder (described below). 
In these calculations only near-neighbour 
hoppings are included, system sizes up to 43,904 sites have been investigated, 
and periodic boundary conditions have been used. The CSTL under consideration 
is quite complicated: there are two formula units per primitive unit cell, 
the conventional unit cell is that of an fcc lattice, and each formula unit 
has two octahedral site Ti atoms. 
That is, each conventional unit cell has 16 sites. For this reason the 
oft-used transfer 
matrix method\cite{Schreiber99} is awkward to implement, and we have instead 
diagonalized 
the Hamiltonian matrix for a lattice with $L\times L\times L$ conventional 
unit cells.

The Hamiltonian for this system is given by
\begin{equation}
\label{eq:Ham_Anderson}
H~=~\sum_{i,\sigma}~\varepsilon_i~n_{i,\sigma}~-~t~\sum_{\langle i,j \rangle,
\sigma}~\Big(c^\dagger_{i,\sigma}~c_{j,\sigma}~+~h.c.\Big)
\end{equation}
where $i,j$ denote the sites of the lattice, $\langle i,j \rangle$ implies
that $i$ and $j$ are near neighbours, and $c_{i,\sigma}$ ($n_{i,\sigma}$) is 
the 
destruction (number) operator for an electron at site $i$ and spin $\sigma$.
The hopping energy is $t$, and the on-site energy at site $i$ is given by 
$\varepsilon_i$.
For a uniform (box) distribution of disorder the on-site energies are being 
chosen at random to be in the range $-W/2$ to $+W/2$, with all energies in 
this range having the same probability. 

Our objective is to analyze this model and determine the energy ranges over 
which the eigenstates are localized and extended -- the trajectory in 
($W$,energy) space defines the mobility edge trajectories. Further, when 
$W$ is increased the upper and lower mobility edges will merge, thus 
identifying the critical disorder; this quantity is denoted by $W_c$ (or, 
more precisely, $W_c/t$ in dimensionless units), 
and we have determined this quantity using several different techniques. 

It seems reasonable that one could be guided, in part, by the extensive 
studies of this transition for the 3d simple cubic lattice; however, there 
are some important differences that make this system considerably more 
difficult to work with. Firstly, the energy spectrum of the simple cubic 
lattice is symmetric about zero energy, and since the upper and lower mobility 
edges merge at the transition indicated by $W_c$, one may
examine an energy window around zero (see, {\em e.g.}, 
Fig. 2 of Ref. \cite{Schreiber99}
for an example density of states for such a problem). Further, since the 
mobility edge is remarkably
flat in the immediate region of $W_c$, one can improve their statistics by 
looking at a reasonably large window of energy around zero (see, {\em e.g.}, 
Fig. 1 of Ref. \cite {Bulka87} and the discussion in Ref. \cite{Hofstetter93}).
However, this is not the case for a CSTL -- the spectrum
is not symmetric about some middle energy (apart from the limit of infinite 
disorder, which is pathological and of no assistance in our numerics), and 
thus the search for the critical disorder (at which the mobility edges merge) 
is more problematic. As an example of this, note that in
studies of the mobility edge trajectories for quantum site percolation models, 
the simple cubic lattice behaves somewhat similarly to the box distribution's 
result discussed above (see Ref. \cite{Soukalis92}), whereas for the CSTL two 
of us and Johnston \cite{Fazileh04} found that the trajectory was not
flat, instead depending strongly on disorder (which for a quantum site 
percolation model is the fraction on unoccupied sites). Indeed, in the results 
presented below, we will show that (i) the mobility edges meet at an energy 
around $-4t$, whereas the middle of the band (for this $W$) is at an energy 
of $-0.5t$, and (ii) the mobility edges do depend on disorder quite strongly,
and the upper and lower mobility edges are not obviously related to one 
another. In part for these reasons, the identification of the mobility edge 
trajectories and $W_c$ is somewhat more challenging for a CSTL.
  
\section{Numerical procedures}

In order to distinguish between localized and extended eigenstates we have
used several techniques. Firstly, we have calculated
the localization length of an eigenstate\cite{Sigeti91} in the form 
\begin{equation}
\label{eq:lambda}
\lambda~=~\sum_i \sum_j \mid\psi_i\mid^2 \mid\psi_j\mid^2 d(i,j)
\end{equation}
where $\psi_i$ is the probability amplitude for the eigenstate,
of a given fixed energy, at site $i$, and $d(i,j)$ is the Euclidean distance 
between lattice sites $i$ and $j$. Then, this quantity is averaged for all
energy eigenstates in a chosen energy range. The ratio of this parameter 
to the identically defined localization length of an eigenstate with constant 
amplitude over the entire lattice, the latter denoted by $\lambda_0$, is
referred to as the relative localization length; the ratio 
is thus denoted by $\lambda/\lambda_0$. 
The behaviour of this relative localization length as a function of increasing 
lattice size (for an specific energy range in the spectrum of the system) 
indicates whether these eigenstates, in the thermodynamic limit, will have 
localized or delocalized character. That is, if this parameter
increases with the system size (having a limiting ratio of one), those states 
are extended, while if this quantity decreases with system size those 
eigenstates are localized. This statistic was employed in the identification
of the mobility edges and metal-to-insulator transition in a quantum site 
percolation model of a CSTL.\cite{Fazileh04,FazilehPhD}

The above-described statistics require the determination of the eigenvectors 
of the Hamiltonian,
which is numerically more demanding than the determination of the eigenvalues 
alone. The most
commonly employed statistic for analyzing the eigenvalues is so-called level 
statistics.\cite{Wegner80,Evers00} However, we have
found\cite{Fazileh05b}
that the most sensitive indication of, in particular, $W_c/t$ is the 
so-called Dyson-Mehta $\Delta_3$ ``spectral rigidity", which is described in 
detail in a variety of references -- see, {\em e.g.}, 
Refs. \cite{BG75,BG83}.~~ $\Delta_3$ is defined by
\begin{equation}
\label{eq:Delta3}
\Delta_3(K)~=~\Big<~\frac{1}{K}~{\min_{A,B}}~\int_x^{x+K}~[N(\varepsilon)
-A\varepsilon -B]^2~d\varepsilon \Big>_x
\end{equation}
where $N(\varepsilon)$ is the integrated density of states, and $<>_x$ denotes 
an average over different parts of the energy spectrum.
At the critical value of disorder this
statistic will be invariant with system size, whereas above and below this 
value of $W$ it will flow to so-called Poissonian and Gaussian Orthogonal 
Ensemble (hereafter referred to as GOE) limits corresponding to localized and
extended states, respectively. The utility of this statistic can be noted in 
its success in identifying 
$W_c/t$ for a variety of models and lattices -- an example for this model of 
disorder for the isotropic 3d simple cubic lattice is discussed in 
Ref.,\cite{Hofstetter93} and for the anisotropic 3d simple cubic lattice 
in Refs.\cite{Milde97,Milde00} Many details of the algorithm that is employed 
calculating this statistic, including the application of this methodology
to the study of other problems, are given in Refs.\cite{BG75,BG83}

We performed these calculations for system sizes up to 16,000 sites when the 
eigenvectors were required, and for sizes up to 43,904 sites when only the 
eigenvalues were evaluated. In a small number of instances we evaluated the 
eigenvectors for system sizes up to 27,648, and found that our predictions 
based on smaller lattices did not change when these results were included.

\section{Numerical results}

\begin{figure}
\begin{center}
\includegraphics[height=8cm,width=11.5cm]{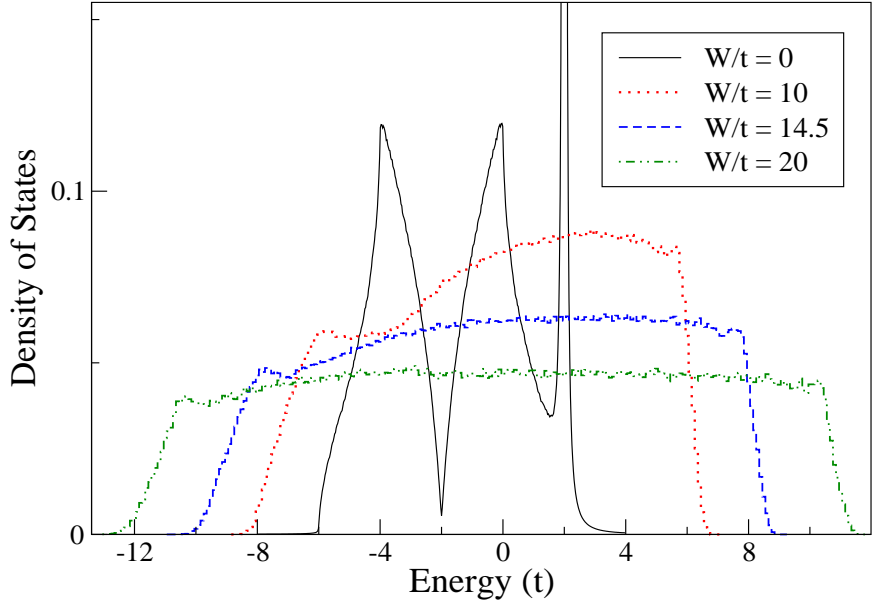}
\caption{\label{fig:DOS_ACSTL} [Colour online]
The density of states for a tight-binding Hamiltonian on a 21,296 site
CSTL with periodic boundary conditions for near neighbour hopping only, 
and with a box distribution of on-site energies of width $W$. For $W\not=0$ 
these data are averages over 50 different realizations of disorder, and the 
final results are found using a histogram method with a bin width of 
$\delta = 0.1t$. The chosen values of $W$ correspond to below, at, and above 
the critical disorder $W_c/t = 14.5$. 
}
\end{center}
\end{figure}

We constructed the complete Anderson disordered Hamiltonians for CSTLs with 
different realizations of box disorder (random diagonal elements between 
$-W/2$ and $W/2$), using periodic boundary conditions, and these have been 
diagonalized. In Fig.~\ref{fig:DOS_ACSTL} we show some representative 
density of states (DOS), including the same quantity for an ordered lattice. 
The data for disordered systems correspond to a system size of 21,296 sites 
averaged over 50 realizations of disorder. The disorder strengths correspond 
to below, equal to, and above the critical disorder (discussed below). 
The evolution of the DOS from the ordered to the disordered systems
show that the zero of the DOS at an energy of $-2t$ (for the ordered lattice) 
becomes a small dip at much lower energies, 
but this dip is not associated with the lower mobility edge -- to be concrete, 
at the critical disorder this dip is present but the upper and lower mobility 
edges have merged (as we show below, they merge at an energy around $-4t$).
At high disorder, as expected, the DOS is approximately symmetric around an 
energy of zero. Lastly, the band edges are qualitatively similar to those 
predicted in analytical calculations.\cite{Wegner80}

Clearly, when the disorder strength is not large (relative to $W_c$) the 
density of states is not symmetric about some ``middle" energy, and to 
clarify previous comments we note that this necessitates that a search for 
the mobility edges must include a broad band of energies in the spectrum. 
In fact, as we discuss below, we find that the mobility edges merge at an 
energy around $-4t$, which is not that close to the middle of the band for 
$W\sim W_c$ (as seen in the figure, this middle is around $-0.5t$). Only in 
the very large $W/t$ limit does the DOS become approximately symmetric about 
zero energy.

The relative localization length, $\lambda/\lambda_0$, was calculated for 
system sizes of $N=$ 128 to 16,000 sites,
and these results were averaged over sufficient
realizations to obtain converged statistics. Typically, this required an average
over a total of roughly 5000 eigenstates in an energy range of $\Delta E = 1.0t$
for any of these system. We found that our numerics displayed 
the desired self averaging over complexions for the larger lattices.

\begin{figure}
\begin{center}
\includegraphics[height=9.5cm,width=12.0cm]{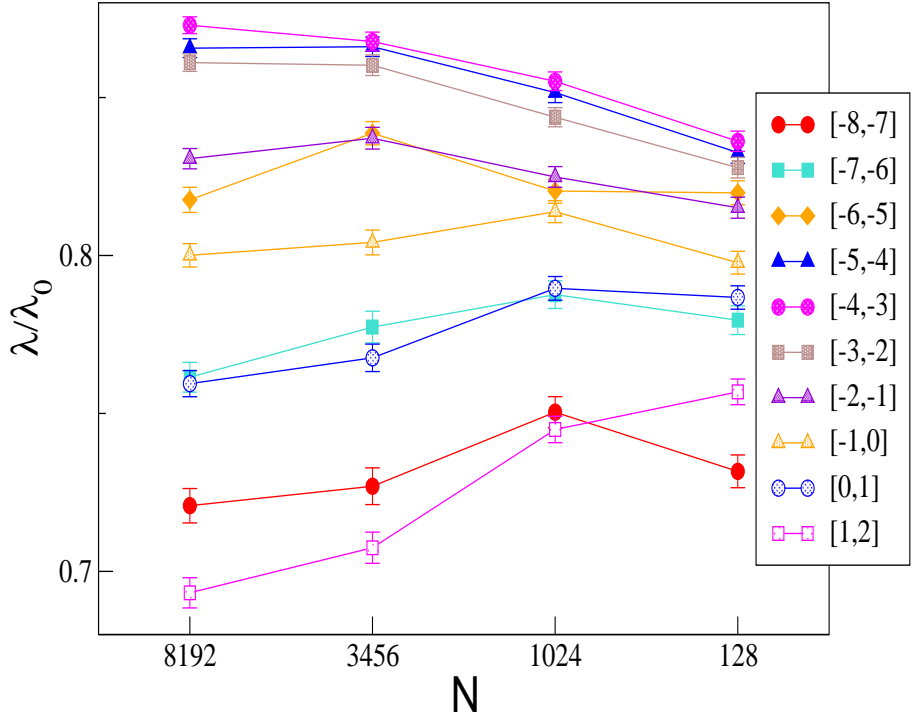}
\caption{\label{fig:W=13.5} [Colour online] 
The variation of the relative localization length for lattice sizes from 128 
to 8192 sites of CSTLs with periodic boundary conditions, as a function of 
different energy ranges, for $W/t = 13.5$. 
}
\end{center}
\end{figure}

A plot displaying typical data is shown in Fig.~\ref{fig:W=13.5} for
a system sizes up of 8192 sites -- this 
corresponds to a $W/t$ ratio that is roughly 7\% below the critical disorder 
(see below). For energy bins that do not include any extended states, the 
relative localization decreases quite strongly as $N$ is increased, as is 
seen above ($E\in[1,2]$) and below ($E\in[-8,-7]$) the extended state energy 
range -- these are states in the upper and lower mobility tails,
respectively. As one approaches the extended state region from above 
($E\in[0,1]$) and below ($E\in[-7,-6]$) the decrease of the relative 
localization length with system size is less pronounced, and the overall 
magnitude of this quantity is increased; these states are also in the mobility 
tails, but are closer in energies to the mobility edges. However, for states 
inside the energy range $E\in[-5,-2]$ we see unambiguously that the relative 
localization length grows as the system size is increased, and these states
correspond to extended states. In the intervening energy ranges ($E\in[-2,0]$ 
and $[-6,-5]$), the lattice sizes shown are too small to clearly 
identify the extended {\em vs.} localized nature of the eigenstates.

\begin{figure}
\begin{center}
\includegraphics[height=16cm,width=12cm]{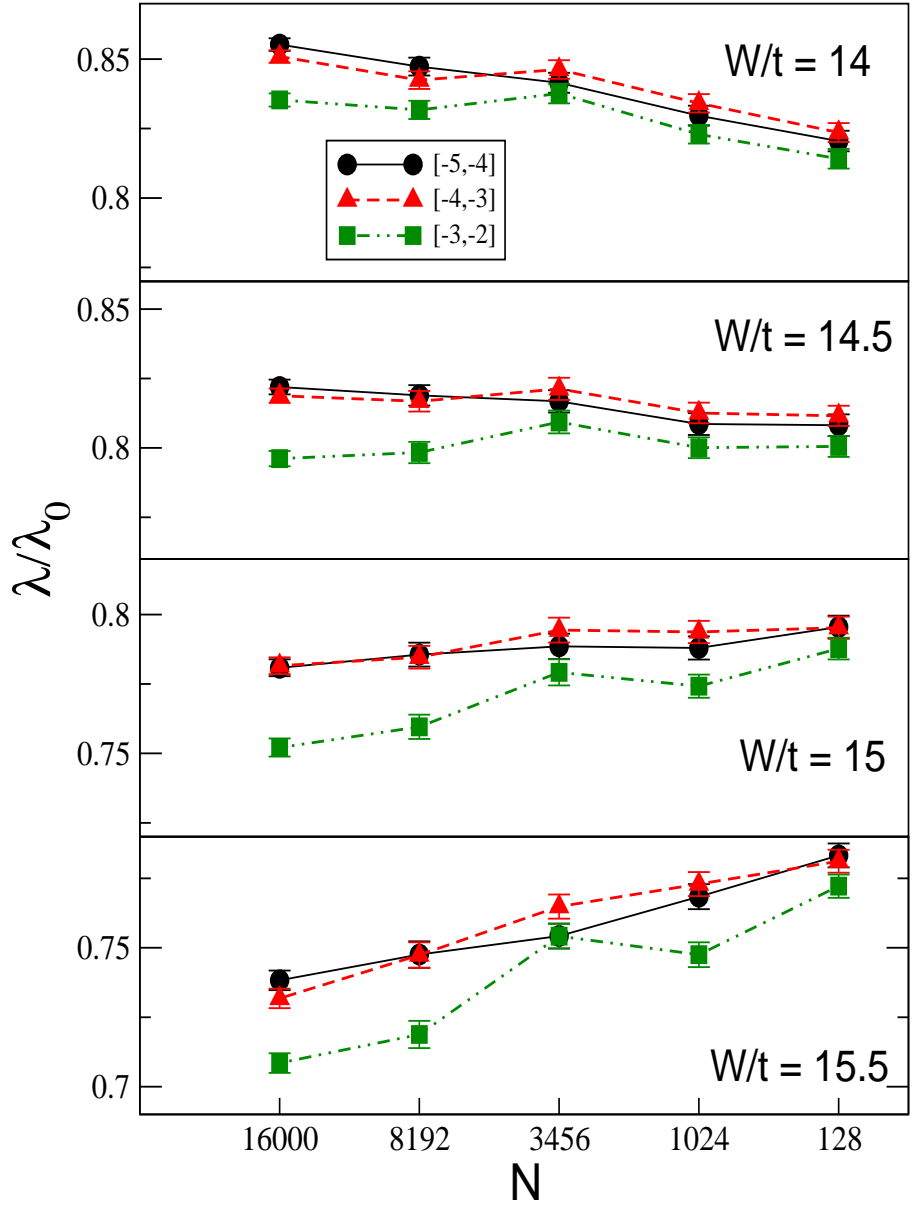}
\caption{\label{fig:W=14to15} [Colour online] Variation
of the relative localization length for systems from $N=$ 128 to 16,000 sites
for $W/t =$ 14, 14.5, 15 and 15.5. Note that the vertical scales are identical,
thus demonstrating the increased (negative) slope of the relative localization
length as N and $W/t$ are increased.
}
\end{center}
\end{figure}

Based, in part, on these results we examined lattice sizes up to 16,000 sites 
in the energy range $E\in[-5,-2]$ with increasing $W/t$ to determine when 
states in this energy region became localized, and our results are shown in 
Fig.~\ref{fig:W=14to15} (note that
the vertical scales are identical in the four frames). For
all energy ranges one sees a clear progression from extended to localized
behaviour with increasing $W/t$, indicated by both the scaling with system 
size and the overall magnitude of the relative localization length. 
For $W/t = 14$, states in the energy range 
$E\in[-5,-2]$ are extended, while for $W/t = 15.5$ these states are localized. 
Also, this data is consistent with states in the energy range $E\in[-3,-2]$ 
being localized for $W/t=15$. This then implies that the mobility edges merge 
at an energy around $-4t$, and, further, this data thus allows us to 
identify a clear upper bound of $W_c/t < 15.5$. Further, recalling our 
earlier statements regarding the scale invariance of the spectral rigidity 
for $W=W_c$, if we also note the 
approximate independence of system size of the relative localization length 
for $W/t = 14.5$ and 15, we expect that $W_c/t$ will be in this range.

In order to identify the mobility edge trajectories it is advantageous to first
have a reliable value for $W_c/t$, but from the above-described relative 
localization length data we cannot be more accurate than 
$W_c/t \sim 14.5 - 15$. Instead, we have found that a determination of both 
the mobility-edge trajectories and the critical disorder may be obtained using 
the scaling of the $\Delta_3$ ``spectral rigidity" of Eq.~(\ref{eq:Delta3}).
To be specific, if this quantity scales to the Poissonian limit of 
uncorrelated energy levels as the size of the system is increased, then the 
states in that energy range are localized, and if this quantity scales to the 
GOE limit the states are extended.\cite{Fazileh05b}

\begin{figure}
\begin{center}
\includegraphics[height=10cm,width=12cm]{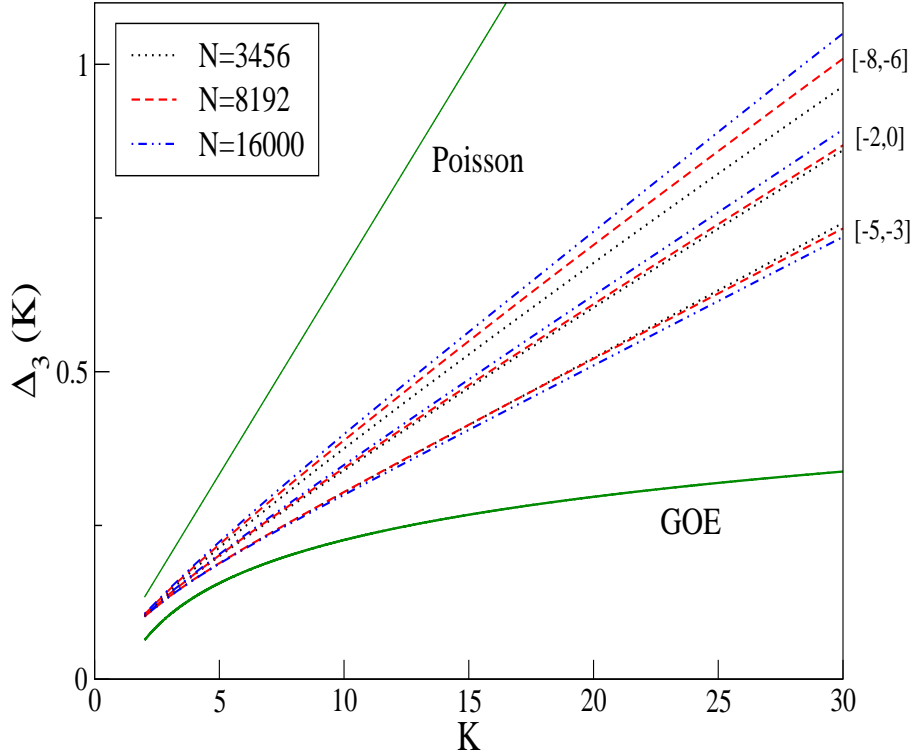}
\caption{\label{fig:Delta3_Weq14_main} [Colour online] The Dyson-Mehta 
spectral rigidity $\Delta_3$
plotted for different energy ranges and system sizes, for $W/t = 14$. 
In the range $E\in[-5,-3]$ these data scale towards the GOE statistics 
(lower curve) or correlated energy levels, whereas
for both $E\in[-5,-3]$ and $E\in[-5,-3]$ these data scale towards the 
uncorrelated result of the
Poissonian limit (upper curve). For clarity we only show three system sizes 
-- other sizes obey the
scaling inferred from the shown data. The two limiting curves correspond to 
(uncorrelated) Poissonian statistics and (correlated) 
Gaussian-orthogonal-ensemble statistics.
}
\end{center}
\end{figure}

We have studied systems having up to 27,648 sites, which is close to the 
number of sites in a 30$^3$ 
simple cubic lattice, and have ensured that for all system sizes a sufficient 
number
of realizations are used such that the spectral rigidity was converged for a 
given $W/t$ and a given system size; sometimes this required that about 
2,000,000 eigenvalues were collected. (For a small number of $W/t$ we have 
found a smaller number of eigenvalues for system sizes up to 43,904, and these 
results are consistent with the data that we present below.) 
For $W/t = 14$ example data is shown in Fig.~\ref{fig:Delta3_Weq14_main}. 
In the energy range of $E\in[-5,-3]$ that was identified 
through the use of the relative localization length (as corresponding to 
extended states as the mobility edges approach one another) $\Delta_3$ moves 
towards GOE statistics, whereas in the $E\in[-8,-6]$ and $[-2,0]$ ranges it 
moves towards the Poissonian limit. 
Further, in the energy ranges $E\in[-3,-2]$ and [-6,-5] we find that there is 
no clear scaling with an increase in system size -- that is, the spectral 
rigidity oscillates with system size in these two energy ranges indicating
the presence of both localized and extended states, and thus one can clearly 
identify that a mobility edge is to be found in this range. 

\begin{figure}
\begin{center}
\includegraphics[height=9.5cm,width=13cm]{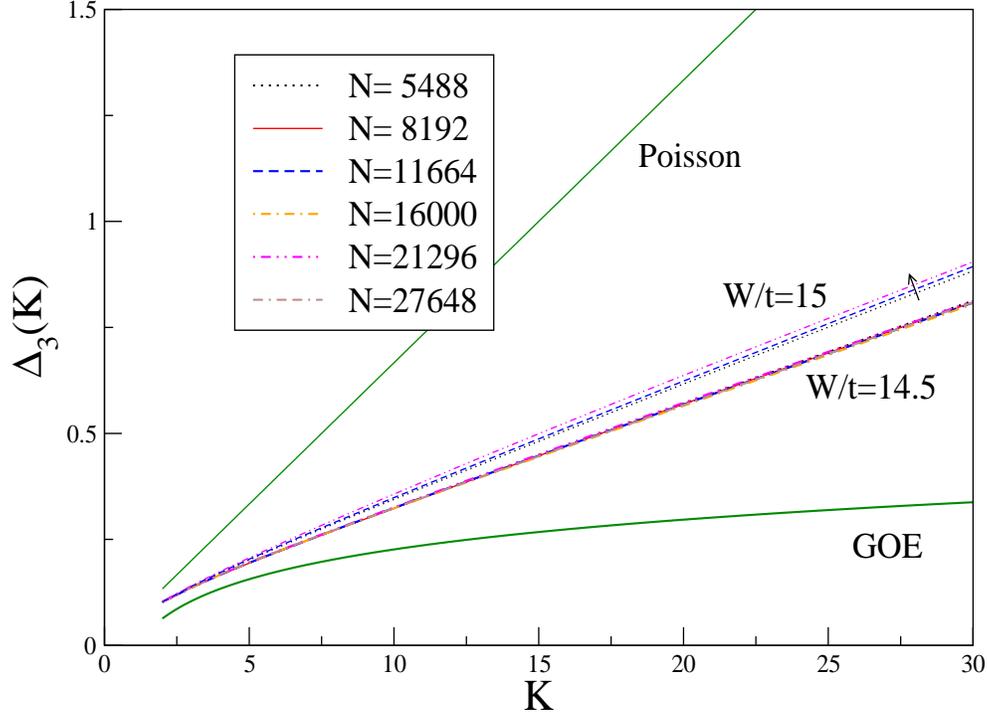}
\caption{\label{fig:Delta3_Wc} [Colour online] 
Scale invariance of the spectral rigidity showing that for the values of 
$W/t = 14.5$ and 15, over
the energy window $E\in[-5,-3]$, this quantity is independent of system size 
for $W/t=14.5$, and scales towards
the Poissonian uncorrelated limit for $W/t=15$. We note for comparison
that these lattice sizes are approximately equivalent to 18, 20, 23, 25, 28, 
and 30 cubed simple cubic lattices.
The Poisson/GOE limits are discussed in a previous caption. The arrow shows 
the direction that the curves
are moving, for $W/t=15$, with increased system sizes (for clarity we only 
show 3 system sizes for $W/t=15$).
}
\end{center}
\end{figure}

As mentioned above, we can also use numerical results for the spectral 
rigidity to determine an accurate value of $W_c/t$. That is, if we examine 
all energy ranges around $E\in[-5,-3]$ and find a scaling of $\Delta_3$ 
to the Poissonian limit for large system sizes, then these states are 
localized and such values of $W_c/t$ are above the critical disorder strength.
Further, we can use the scale invariance of the spectral rigidity to precisely
identify $W_c/t$, and our results are shown in Fig.~\ref{fig:Delta3_Wc}. 
We find that for $W/t = 14.5$ this quantity is independent of system size. 
We examined a mesh of $\Delta W/t=0.25$ for $W/t$ around 14.5, but found that 
for precisely $W/t = 14.5$ the scale invariance was the most robust --
the curves for all system sizes are very nearly coincident, and most 
importantly fluctuate in a very small width about an average (as opposed to 
moving towards the Poissonian or GOE limits).
So, our final number for the critical disorder of a 
box distribution of disorder in an Anderson model for CSTLs is 
$W_c/t = 14.5 \pm 0.25$. For $W/t=15$ this figure shows that the states in 
this energy range are scaling towards the uncorrelated limit,
thus confirming that such disorder is in excess of the critical disorder.
 
\begin{figure}
\begin{center}
\includegraphics[height=10cm,width=13cm]{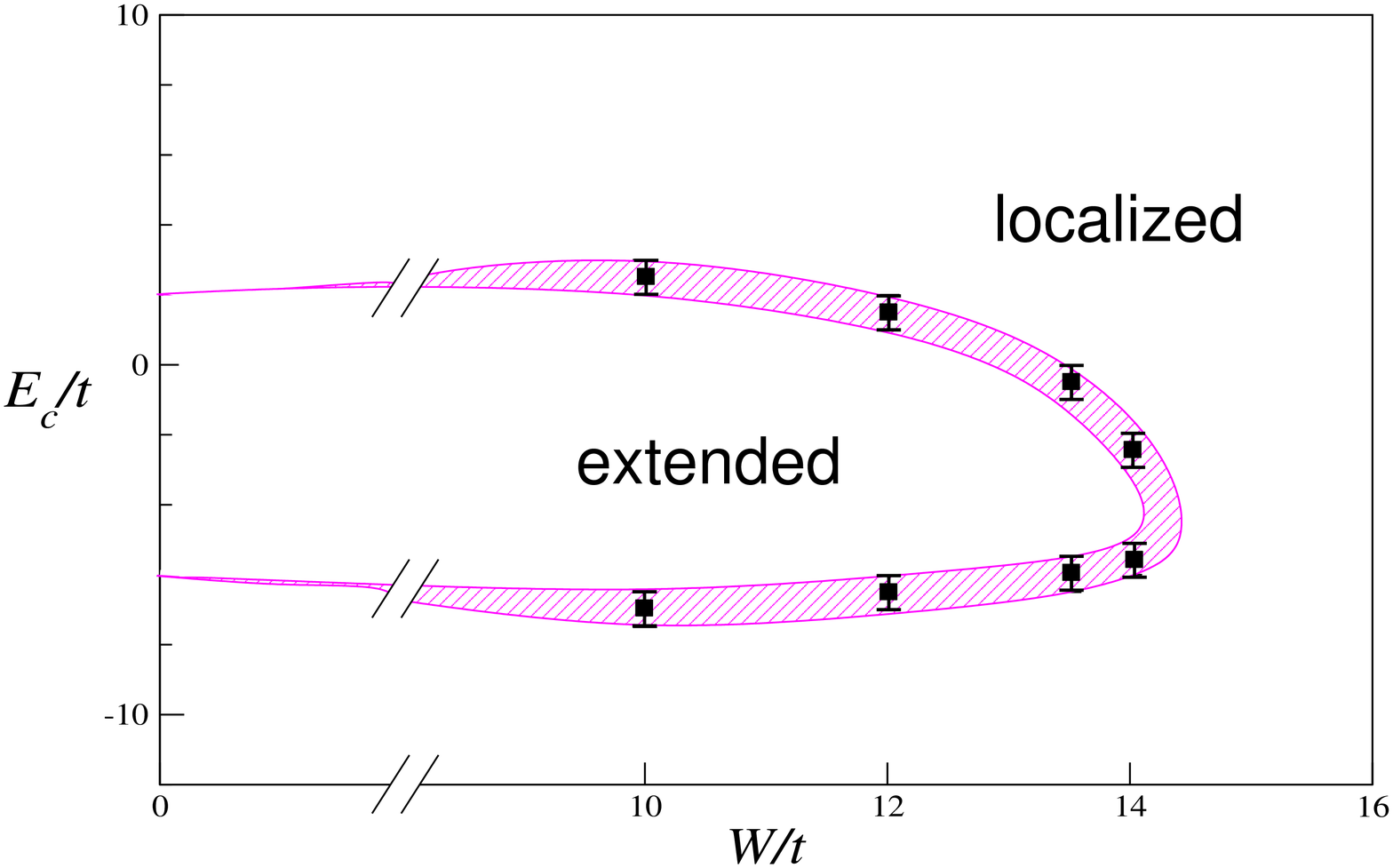}
\caption{\label{fig:medge_box_CST} [Colour online] 
The mobility-edge trajectories (estimates bounded by hatched region) separating
localized and extended states on a CSTL for a box distribution of on-site 
energies of width $W$, with our data points (solid squares) and error bars 
interpolated by the boundaries of the hatched region.
}
\end{center}
\end{figure}

Using such a combined relative localization length/spectral rigidity approach, 
we have identified the mobility edges for a variety of $W/t$, and our estimates
for these energies as a function of 
disorder are shown in Fig.~\ref{fig:medge_box_CST}.
As discussed previously, it is interesting to note that the energy at which 
the mobility edges merge at $W_c/t$
is not in the middle of the density of states (also see 
Fig.~\ref{fig:DOS_ACSTL}); further, unlike the simple cubic
lattice, our data for the the mobility edges show that they do not rapidly 
coalesce as $W_c/t$ is approached from below. Instead, the upper mobility 
edge approaches the energy at which the two edges merge much more gradually 
-- for comparison, see Fig.~1 of reference.\cite{MacKinnon83}

One further comparison of these results to previously published data for 
three-dimensional simple cubic lattices involves the magnitude of $W_c/t$ 
-- note the well accepted result for the critical value of $W_c/t$ for the 
simple cubic lattice, \textit{viz.} 16.5.
\cite{MacKinnon83,Schreiber95,Slevin97,Zharekeshev97,Ohtsuki00,Slevin01}
At this time we do not have a reliable explanation for the decreased value
of $W_c/t$ of the CSTL, something that can only be realized with detailed 
analytical work.
However, if one follows the logic that $W_c$ should be close to the 
noninteracting bandwidth, which is 12$t$ for the simple cubic lattice and 8$t$ 
for the CSTL, and thus re-expresses $W_c$ in units of the bandwidth (which 
here we label by $B$), then
$W_c/B~\sim~1.4$ for the simple cubic lattice, and $W_c/B~\sim~1.8$ for 
the CSTL, and thus
one sees that in these scaled units the critical disorder is in fact larger in 
CSTLs than in a simple cubic lattice. One can speculate that the increase of 
$W_c/B$ follows from the various near neighbours in these two lattices -- we 
have listed the near neighbours up to the 10th ``shell" in 
Table~\ref{table:nn_d}. From this one sees that in the 3rd and 4th shells the 
CSTL has many more neighbouring sites on which to form delocalized states than 
does the simple cubic lattice, so perhaps this difference between these 
lattices is partly responsible for the large increase of $W_c/B$ in the CSTL
in comparison to the simple cubic lattice.

\begin{table}
\caption{\label{table:nn_d}\\ A comparison of the number of nearest neighbours 
(n.n.) for simple cubic and CSTLs.}
\begin{ruledtabular}
\begin{tabular}{@{}ccc}
Near Neighbours   & Simple Cubic Lattice & CST Lattice  \\
1st   & 6  & 6  \\
2nd   & 12 & 12 \\
3rd   & 8  & 12 \\
4th   & 6  & 12 \\
5th   & 24 & 24 \\
6th   & 24 & 6  \\
7th   & 12 & 18 \\
8th   & 30 & 12 \\
9th   & 24 & 24 \\
10th  & 24 & 36 \\
\end{tabular}
\end{ruledtabular}
\end{table}

\section{Application to L\lowercase{i}T\lowercase{i}$_2$O$_4$}

We have presented comprehensive numerical results for a box distribution of 
random on-site energies for CSTL having near-neighbour hopping only. We have 
used the relative localization length and spectral rigidity to identify the 
mobility-edge trajectories as a function of the width of the box distribution, 
and have determined that $W_c/t \sim 14.5$ is the critical disorder 
at which the upper and lower mobility edges merge. These results aid us in 
understanding results presented below for disordered LiTi$_2$O$_4$.\cite{LiAlTiO}

As was discussed in the introduction, our work on this problem was motivated 
by our interest in the metal-to-insulator transition undergone by 
Li$_{1+x}$T$_{2-x-y}$Al$_y$O$_4$. One way to go beyond the quantum site 
percolation calculations discussed in reference\cite{Fazileh04} is to include 
the change of the on-site energies due to the impurity potentials,
and we have determined the distribution of on-site energies produced by these 
impurity potentials (using different models of screening). A detailed analysis 
of the resulting distributions, independent of the model of screening used, 
indicated that almost all of these energies were between approximately 
$\pm t$, which, if represented by an Anderson model with a box distribution
of on-site energies, would correspond to $W/t =2$, well below $W_c/t = 14.5$ 
for CSTL. 

\begin{figure}[t]
\begin{center}
\includegraphics[height=7.5cm,width=9.5cm]{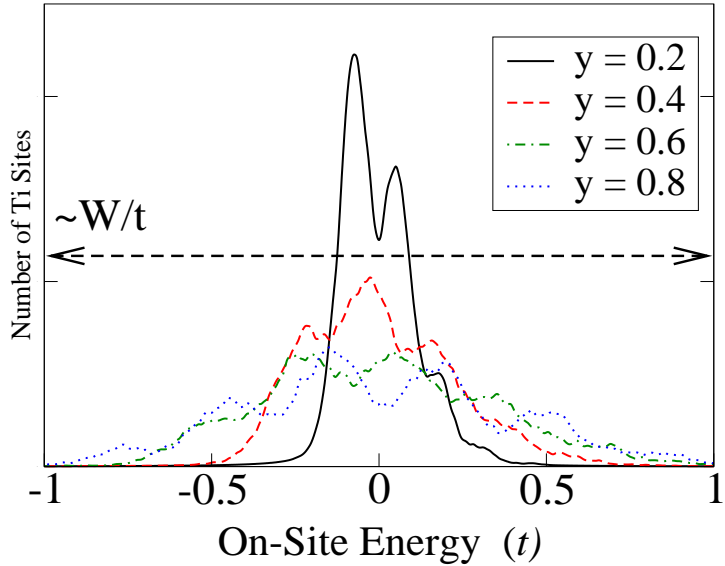}
\caption{\label{fig:SiteNrgs} [Colour online]
The distribution of the on-site energies of the Ti sites in a disordered lattice
as a function of Al concentration $y$. These energies are found in a Lindhard
model of screening of the impurity potentials.
}
\end{center}
\end{figure}

To determine this estimate, a particular complexion of disorder is considered in which
the Al ions are placed randomly on the octahedral sites of the spinel structure
(the Ti sites of the ordered crystal) according to $y$, which determines the
concentration of Al impurities, and then the screened Coulombic potential from
all Al sites is summed up for each of the (non-disordered) Ti sites -- these energies
thus approximate the (relative) on-site energies of the Ti sites in the disordered lattice. 
This procedure is repeated for many different realizations of disorder.
An example plot of such a distribution of  on-site energies is given in Fig.~\ref{fig:SiteNrgs} 
for the case of Lindhard screening, which is introduced in the usual manner;\cite{AMtext}
very similar distributions are found for a Thomas-Fermi model of screening.
Thus, we do not expect the inclusion of this ``additional disorder" to produce a 
substantial change of the results found before,\cite{Fazileh04} for which the only 
disorder effects that were included arose from quantum site percolation.

\begin{figure}
\begin{center}
\includegraphics[height=9.5cm,width=13cm]{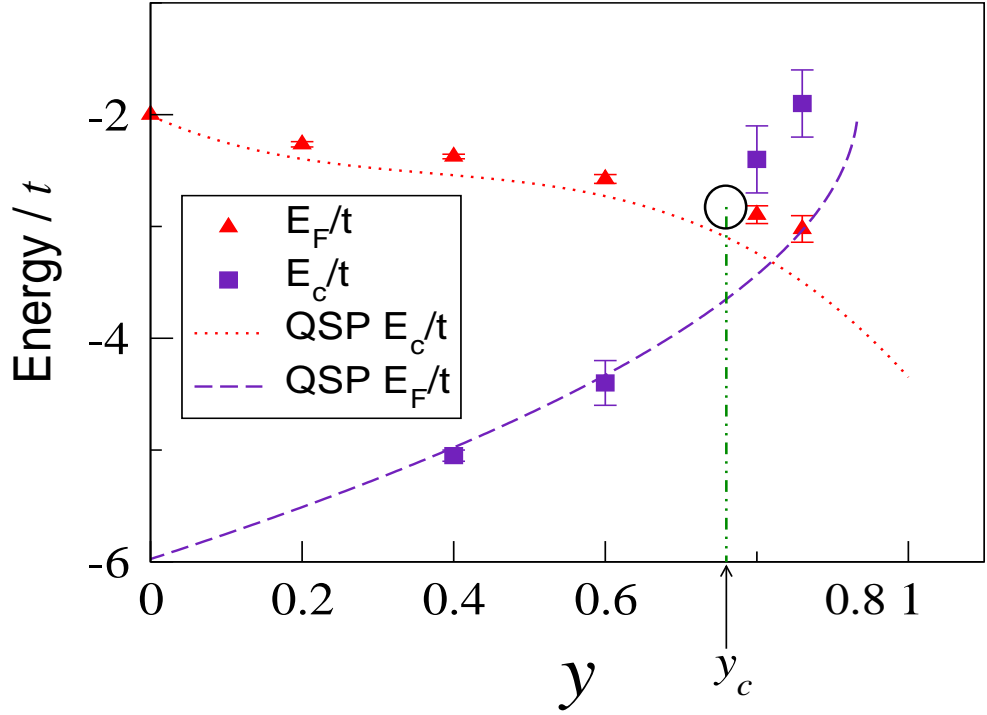}
\caption{\label{fig:LiTiO_with_imps} [Colour online]
The numerically determined phase diagram for the identification of
the critical doping concentrations of the metal-to-insulator transition
in LiAl$_y$Ti$_{2-y}$O$_4$ after including the on-site
energies from the screened Coulomb potentials of the doping Al$^{3+}$ ions;
the Lindhard approximation is used for screening potentials.
The data shown are the Fermi energies ($E_F$) and mobility edges ($E_c$) as a
function of doping for \LiAlTiO~system.
Broken lines (dotted and dashed) are estimates of the Fermi energies and the mobility edges
from our previous calculations of the quantum site percolation (QSP) model only
-- see reference.\cite{Fazileh04} The open black circle is the new estimate of 
the $Al$ concentration at which one would find a metal-to-insulator transition, in
a model that includes the effects of both quantum site percolation and the inhomogeneous
distribution of on-site energies caused by the $Al$ impurity potential.
}
\end{center}
\end{figure}

Indeed, that is what we concluded when we repeated the calculations described 
in reference \cite{Fazileh04} but now adding in the varying on-site energies 
generated by the Al impurity potentials --  our results for LiAl$_y$Ti$_{2-y}$O$_4$ 
when a Lindhard model of screening is used are shown in Fig.~\ref{fig:LiTiO_with_imps}. 
We find that the critical value of Al doping that leads to 
a metal-to-insulator transition is reduced from $y_c \approx 0.82$ for
a model that includes quantum site percolation only to $y_c \approx 0.78$ when 
the effects of impurity potentials are also included. That is, the role played 
by the impurity potential disorder is minimal -- the lower mobility-edge energy, 
for a given disorder, is increased only moderately in a model that includes
the effects of the Al impurity ion potentials.

\section{Conclusions}

We have determined the mobility-edge trajectories and the critical disorder for
corner-sharing tetrahedral lattices, lattices that are common in studies of 
fully frustrated magnetic systems, as well as the sublattice of octahedral sites 
of a normal spinel structure. We have examined the metal-to-insulator transition
of LiAl$_y$Ti$_{2-y}$O$_4$ and determined that a quantum site percolation model
plus Anderson-like on-site disorder produced by impurity potentials leads to
a critical doping of $y_c \approx 0.78$, only 5\% less than one estimates if
Anderson disorder is ignored and only quantum site percolation is studied. 
Since $y_c \approx 0.33$ is the experimental
value,\cite{LiAlTiO} indirectly this result supports the hypothesis that 
something beyond one-electron physics is required to explain this transition, {\em e.g.}
strong electronic correlations. This possibility will be addressed in a future
publication.\cite{Fazileh05}

\begin{acknowledgments}
We thank George Sawatzky, Bill Atkinson, and Gene Golub for helpful 
comments, and Pakwo Leung for making available some of his computing resources.
This work was supported in part by the NSERC of Canada, and the Ontario 
Graduate Scholarship in Science and Technology programme.
\end{acknowledgments}


\newpage

\begin{thebibliography}{1}

\bibitem{CJP01} See, \textit{e.g.,} the conference proceedings in 
Can. J. Phys.  {\bf 79}, 11/12 (2001).
                                                                                
\bibitem{PWA56} P. W. Anderson, Phys. Rev. {\bf 102}, 1492 (1958).

\bibitem{pyroM} S. T. Bramwell and M. J. P. Gingras, Science {\bf 294}, 
1495 (2001).

\bibitem{pyroSC} H. Aoki, J. Phys.: Cond. Mat. {\bf 16}, V1 (2004).

\bibitem{Barth32} T. F. W. Barth and E. Posnhak, Z. Krist. {\bf 82}, 325 (1932).

\bibitem{Goodenough60} J. B. Goodenough, Phys. Rev. {\bf 117}, 1442 (1960).

\bibitem{dcj97} S. Kondo, D. C. Johnston, C. A. Swenson, F. Borsa, 
A. V. Mahajan, L. L. Miller, T. Gu, A. I. Goldman, M. B. Maple, 
D. A. Gajewski, E. J. Freeman, N. R. Dilley, R. P. Dickey, J. Merrin, 
K. Kojima, G. M. Luke, Y. J. Uemura, O. Chmaissem, and J. D. Jorgensen, 
Phys. Rev. Lett. {\bf 78}, 3729 (1997).

\bibitem{dcj76} D. C. Johnston, J. Low. Temp. Phys. {\bf 25}, 145 (1976).

\bibitem{muller96} K. A. M\"uller in {\it Proceedings of the 10$^{th}$ 
Anniversary HTS Workshop on Physics, Materials and Applications} ed 
B. Batlogg {\it et al.} (World Scientific: Singapore) page 3 (1996).

\bibitem{Fazileh04} F. Fazileh, R. J. Gooding and D. C. Johnston, Phys. Rev. B
{\bf 69}, 104503 (2004).

\bibitem{Fazileh05} F. Fazileh, R. J. Gooding, W. A. Atkinson, and D. C. Johnston, 
to appear in Phys. Rev. Lett. - Feb. 3, 2006.

\bibitem{FazilehPhD} F. Fazileh, PhD Thesis, Queen's University (2005) -- unpublished.

\bibitem{fuji01} S. Fujimoto, Phys. Rev. B {\bf 64}, 085102 (2001).

\bibitem{fuji02a} S. Fujimoto, Phys. Rev. B {\bf 65}, 155108 (2002).

\bibitem{fuji02b} S. Fujimoto, Phys. Rev. Lett. {\bf 89}, 226402 (2002).

\bibitem{fuji03} S. Fujimoto, Phys. Rev. B {\bf 67}, 235102 (2003).

\bibitem{chen02} C. Chen, Phys. Lett. A {\bf 303}, 81 (2002).

\bibitem{imai02} Y. Imai and N. Kawakami, Phys. Rev. B {\bf 65}, 233103 (2002).

\bibitem{Schreiber99} M. Schreiber, F. Milde and R. A. R\"omer, Comp. Phys. 
Comm. {\bf 121-122}, 517 (1999).

\bibitem{Bulka87} B. Bulka, M. Schreiber and B. Kramer, Z. Phys. B {\bf 66}, 21 (1987).

\bibitem{Hofstetter93} E. Hofstetter and M. Schreiber, Phys. Rev. B {\bf 48}, 16979 (1993).

\bibitem{Soukalis92} C. M. Soukalis, Q. Li and  G. S. Grest, Phys. Rev. B {\bf 45}, 7724 (1992).

\bibitem{Sigeti91} D. E. Sigeti, X. Zhang, M. S. Friedrichs, and R. A. Friesner, Phys. Rev. B {\bf 44}, 614 (1991).

\bibitem{Wegner80} F. Wegner, Z. Phys. B {\bf 36}, 209 (1980).

\bibitem{Evers00} F. Evers and A. D. Mirlin, Phys. Rev. Lett. {\bf 84}, 
3690 (2000).

\bibitem{Fazileh05b} We have also completed extensive analysis of this system 
using the 2nd moment of the inverse participation ratio, and these data 
further confirm the results presented in this manuscript.

\bibitem{BG75} O. Bohigas and M. J. Giannoni, Annals of Physics {\bf 89}, 393 (1975).

\bibitem{BG83} O. Bohigas and M. J. Giannoni, {\it Mathematical and 
Computational Methods in Nuclear Physics} ed. J. S. Dehesa {\it et al.} 
(Springer-Verlag, Berlin) page 1 (1983).

\bibitem{Milde97} F. Milde, R. A. R\"omer and M. Schreiber, Phys. Rev. B 
{\bf 55}, 9463 (1997).

\bibitem{Milde00} F. Milde, R. A. R\"omer and M. Schreiber, Phys. Rev. B 
{\bf 61}, 6028 (2000).

\bibitem{MacKinnon83} A. MacKinnon and B. Kramer, Z. Phys. B {\bf 53}, 1 
(1983).

\bibitem{Schreiber95} H. Grussbach and M. Schreiber, Phys. Rev. B {\bf 51}, 
R663 (1995).

\bibitem{Slevin97} K. Slevin and T. Ohtsuki, Phys. Rev. Lett. {\bf 78}, 4083 
(1997).

\bibitem{Zharekeshev97} I. K. Zharekeshev and B. Kramer, Phys. Rev. Lett. 
{\bf 79}, 717 (1997).

\bibitem{Ohtsuki00} T. Kawarabayashi, T. Ohtsuki and K. Slevin, Physica B 
{\bf 284-288}, 1549 (2000).

\bibitem{Slevin01} K. Slevin and T. Ohtsuki, Phys. Rev. B {\bf 63}, 045108 
(2001).

\bibitem{LiAlTiO} P. M. Lambert, P. P. Edwards and M. R. Harrison, 
J. Sol. St. Chem. {\bf 89}, 345 (1990).

\bibitem{AMtext} N.W. Ashcroft and N.D. Mermin, \textit{Solid State Physics}
(Saunders College, Philadelphia, 1976).

\end{thebibliography}
\end{document}